\documentclass[twocolumn,preprintnumbers,amsmath,amssymb]{revtex4}
\usepackage{graphicx}
\usepackage{dcolumn}
\usepackage{bm}

\begin{document}
\title{\bf Strain engineered  graphene using a nanostructured substrate: I Deformations}
\author{M. Neek-Amal$^{1,2}$, and F. M. Peeters$^2$ }
\affiliation{$^1$Department of Physics, Shahid Rajaee Teacher Training University,
Lavizan, Tehran 16785-136, Iran.\\$^2$Departement Fysica,
Universiteit Antwerpen, Groenenborgerlaan 171, B-2020 Antwerpen,
Belgium.}
\date{\today}
\begin{abstract}
Using atomistic simulations we investigate the morphological
properties of graphene deposited on top of a nanostructured
substrate. Sinusoidally corrugated surfaces, steps, elongated
trenches, one dimensional and cubic barriers, spherical bubbles,
Gaussian bump and Gaussian
 depression are considered as  support structures for graphene.
The graphene-substrate interaction is governed by van der Waals
forces and the profile of the graphene layer is determined by
minimizing the energy using molecular dynamics simulations. Based on
the obtained optimum configurations, we found that: (i) for graphene
placed over  sinusoidally corrugated substrates with corrugation
wave lengths longer than 2\,nm, the graphene sheet follows the
substrate pattern while for supported graphene it is always
suspended across the peaks of the substrate, (ii) the conformation
of graphene to the substrate topography is enhanced when increasing
the energy parameter in the van der Waals model, (iii) the adhesion
of graphene into the trenches depends on the width of the trench and
on graphene's orientation, i.e. in contrast to  a small width
(3\,nm) nanoribbon with armchair edges, the one with zig-zag edges
follows the substrate profile, (iv) atomic scale graphene follows a
Gaussian bump substrate but not  the substrate with a Gaussian
depression, and (v) the adhesion energy due to van der Waals
interaction varies in the range [0.1-0.4] J/m$^2$.

\end{abstract}
\maketitle

\section{Introduction}
Geometrically structured substrates affect  various properties of
graphene~\cite{geim,novo}, and can prevent the crumpling of graphene
which is typical for free standing graphene without a
support~\cite{nelson}. Before graphene's discovery in 2004, the
study of 2D membranes over corrugated substrates was an important
branch of soft-condensed matter physics with e.g. applications in
biological systems~\cite{langmuir,PREswin}. Recently, particular
attention was focused on the various properties of graphene on top
of a substrate. Substrates can induce corrugations, modify the
electric conductance and deform graphene~\cite{bao,sio2}. The
electrostatic interaction of graphene on a  substrate can be
understood as due to the van der Waals (vdW) interaction of graphene
with the metallic gate below the substrate, the electrostatic forces
between graphene and the  polarized substrate, the water between the
substrate and graphene, and impurities between graphene and the
substrate~\cite{Sabio}. The vdW interaction includes  attractive and
repulsive terms, which are widely used and extensively investigated
in soft matter~\cite{safran}. The usual dispersion interaction for
the attractive part of the vdW interaction (sum of contributions
proportional to $D^{-6}$) must be modified for the two
$\pi$-conjugated systems at distance $D$, e.g.
graphite~\cite{PRL2006}. For the vdW interaction between carbon
nanostructures and Si-C/SiO$_2$ substrates the Lennard-Jones
potential (LJ) is widely used and produced both qualitative and
quantitative acceptable
results~\cite{MD2010,Ref27MD2010,Ref28MD2010,neek2010,PRB2010,ACSnano}.
Ab-initio calculations obtained  vdW energy curves between carbon
nanostructures (those curves are similar curves to the LJ function),
e.g the vdW interaction between hexagonal boron nitride sheets
\cite{PRL2010BN}, methane adsorption on graphene and gas molecule
adsorption in carbon nanotubes were found to be LJ like functions
when using vdW corrected density functional
theory~\cite{nanotechnology2002,surfacescience}.

Recently, experimental measurements on the adhesion energy of
pressurized mono-layer/multi-layer graphene on top of a SiO$_2$
substrate showed that the adhesion energy is ultra strong ($\chi\sim
0.3-0.45$\,J/m$^2$) which is many times larger  than the one
reported for typical micromechanical structures and is of the order
of solid-liquid adhesion energies~\cite{naturenano}. This adhesion
energy is one order of magnitude larger than the upper limit found
for water modified  adhesion between graphene and the substrate.
Kusminskiy~\emph{et al} used $\chi$=2 meV\AA$^{-2}$~ for the pinning
of a tethered membrane (as a model for graphene within continuum
elasticity theory) and found the possible morphology of graphene
over a Gaussian bump and a Gaussian depression~\cite{Piniing}. Their
model includes both bending and stretching energies together with a
constant pinning energy.

Here, as distinct from previous works, we investigate graphene on
top of several nanostructured substrates with different geometrical
deformations. We carried out molecular dynamics simulations at
$T$=300\,K to minimize the energy and find the optimum profile of
the deposited graphene membrane. Sinusoidal substrates with
different wave lengths, elongated trenches, barriers, bubbles,
Gaussian bump and Gaussian depressions are considered as geometrical
distinct examples of nanostructured substrates. We find that in case
of a sinusoidal substrate with short wave length and small energy
parameter in the vdW model (i.e. $\epsilon$), graphene does not
follow the substrate. For graphene on top of the trench, it is found
that  zig-zag graphene falls into the well but arm-chair graphene is
suspended across the trench. The stress distribution shows that the
atoms within the deformed parts are highly stressed. For the
boundary conditions of the examined graphene flakes we considered
both free and supported (i.e. fixed) in-plane boundaries. We found a
significant difference in the obtained graphene profile when on top
of a Gaussian bump or at a Gaussian depression, i.e. the graphene
sheet over a depression with 1\,nm variance and 1\,nm height does
not fall into the depression while it follows a Gaussian bump with
the same size. For a Gaussian bump/depression with larger variance,
graphene follows both substrates. The square barrier (a cube with
1\,nm side) influences graphene such that an unexpected pyramidal
shape is found which surrounds the barrier. We studied the vdW
energy stored between graphene and the nanoscale Gaussian bump by
employing a continuum model for both systems and calculated the
variation of the vdW energy per area as a function of the energy
parameter of the model. We also compared our molecular dynamics
results for the Gaussian bump/depression to those predicted by the
continuum model and found agreement only in case of a large Gaussian
bump with weak interaction , i.e. small $\epsilon$.

This paper is organized as follows. In Sec. II the details of the
atomistic model are presented.  In Sec. III we present the continuum
model for the vdW interaction of graphene and various substrates.
In Sec.~IV we present  results  for various nano-structured
substrates and different boundary conditions. The results are
summarized in Sec. V.

\section{Atomistic model} Classical atomistic molecular dynamics
simulation (MD) is employed to simulate large flakes of graphene
(GE). The second generation of Brenner's bond-order potential is
employed which is able to describe covalent sp$^3$ bond breaking and
the formation of associated changes in the atomic hybridization
within a classical potential~\cite{brenner2002}. The Brenner
potential terms were taken as

\begin{equation}\label{Eqbond}
E_P=\sum_i\sum_{j>i}[ V^R(r_{ij})-B_{ij}V^A(r_{ij}],
\end{equation}
where $E_P$ is the average binding energy, and $V^R$ and $V^A$ are
the repulsive and attractive terms, respectively, where $r_{ij}$ is
the distance between the atoms \emph{i} and \emph{j} (all relevant
parameters in this study are listed in Table I). $B_{ij}$ is called
the bond order factor which includes all many-body effects. $B_{ij}$
depends on the local environment of the bond, i.e. the bond and
torsional angles, the bond lengths and the atomic coordination in
the vicinity of the bond. This feature allows the Brenner potential
to predict correctly the configurations, the hybridization and the
energies for many different hydrocarbon structures.

 The carbon-carbon bond length, $a_0$, is 1.42\,\AA. In our model, the
origin of the $xyz$-Cartesian coordinate system is set at (0,0,0).
Here, the two primitive vectors for the GE sublattices,
$\textbf{a}_1=\sqrt{3}a_0\hat{i}$ and
$\textbf{a}_2=\sqrt{3}/2a_0\hat{i}+3/2a_0\hat{j}$ are the two basic
vectors of GE lattice.

\begin{table*}
\caption{List of all relevant parameters used in  the paper.}
\begin{tabular}{| l | l | c | c |}
  \hline
$l_x,l_y$&The length and width of the simulated graphene membrane\\
$\epsilon,\sigma$&The energy and length parameters in the van der Waals (vdW) potential for the substrate atoms,  Eq.~(\ref{Equr})\\
$\alpha,\beta,m$&Integer numbers defining the power law of repulsive and attractive parts of the vdW potential in Eq.~(\ref{Equr})\\
$\eta^i,\textbf{F}^i$&The stress and the force on atom \emph{i}\\
$\emph{f},E_{vdW}$&Total  free energy (Eq.~(\ref{EqF})) and vdW energy  (Eqs.~(\ref{Eq1}),(\ref{EqC})) stored in graphene when on top of a substrate\\
$m,\Omega$,$\textbf{v}_i$,$\textbf{r}_i$,&Mass, volume, velocity and position of atom \emph{i}\\
$N,M$&Total number of atoms respectively, of the simulated graphene and the substrate\\
$\tau,\kappa$&Surface tension, bending rigidity of the graphene membrane\\
$\chi,T$&The vdW energy per planar area `A' and temperature\\
$h_G(x,y),h_S(x,y)$&Height of the graphene membrane  and the substrate at (x,y)\\
$E_P,B_{ij},V_R,V_A$&Total bond energy of graphen, bond order, repulsive and attractive terms in the Brenner potential, Eq.~(\ref{Eqbond}) \\
$v,\Sigma_S,\Sigma_G$&Hamaker constant, the density of the simulated substrate and graphene membrane, respectively\\
$\lambda,\theta(x)$&The wave length and the step function\\
$R,g(\textbf{r})$&The radius of Gaussian bump and the determinant of the metric tensor\\
$h_0$&The amplitude of the sinusoidal waves or height (depth) of Gaussian bump/bubble/barrier (depression or trench)\\
$h_1,d$&A shift or vertical distance between graphene and the substrate and the width of the trenches/barriers\\
$u(r),r_{min}$&The Lennard-Jones potential energy and its minimum distance\\
$a_0,\textbf{a}_1,\textbf{a}_2$&Carbon-carbon bond length and two basic vectors of the graphene lattice \\
$\ell$&Lattice parameter of the substrate lattice\\
$r_c,r_{0}$&Cutoff distance in vdW interaction, upper limit of the integrals in Eq.~(\ref{EqC})\\
$ r_{ij}$&Distance between a lattice site  of graphene (`i') and a lattice site of the substrate (`j') \\
$\Delta r_{12}$&Distance between a surface element of the graphene membrane  and a surface element of the substrate\\

 \hline
\end{tabular}
\label{table1}
\end{table*}

In order to model the van der Waals (vdW) interaction between GE and
different substrates, we employed the Lennard-Jones (LJ) potential.
The LJ potential describes both the repulsive and attractive parts
of the vdW energy between two atoms which are non-bonded. The LJ
potential is a widely used potential in various
simulations~\cite{MD2010,Ref27MD2010,Ref28MD2010,neek2010,PRB2010,ACSnano}.
For two interacting uncharged particles, we have
\begin{equation}\label{Equr}
u(r)=\epsilon[\alpha(\sigma/r)^{3m}-\beta(\sigma/r)^6],
\end{equation}
 where $r$ is the distance between the two particles,
$\epsilon$ and $\sigma$ are the `\emph{energy parameter}' and  the
`\emph{length parameter}', respectively. Setting the integer numbers
to $\alpha=\beta=m=4$ Eq. (2) gives the 12-6 LJ potential function
and for $\alpha=2,\beta=m=3$ Eq. (2) gives the 9-6 LJ potential
function~\cite{JAPL2010}.
 The minimum of $u(r)$ is
$r_{min}=\sigma(\frac{m\alpha}{2\beta})^{1/(3m-6)}$ which yields the
$2^{1/6}\sigma$ and $\sigma$ for 12-6 LJ and 9-6 LJ potential,
respectively. Therefore, the equilibrium distance is shorter for the
9-6 LJ potential while the minimum of $u(r)$
($u|_{r_{min}}=-\epsilon$) is the same for both cases.  Notice that
for the 12-6 LJ potential both attractive and repulsive terms have
the same weights in $u(r)$, i.e. $\alpha=\beta$.  We will use mostly
the 12-6 LJ potential (in some exceptional cases we use the 9-6 LJ
potential will be mentioned expliciltly).

To model the interaction between two different types of atoms such
as the carbon atom (C) and a substrate atom (S), we adjust the LJ
parameters using the equations $\epsilon_T\,=\, \sqrt{\epsilon_C
\epsilon}$ and $\sigma_T\,=(\sigma_{C}+\sigma)/2$. For carbon we use
the parameters  $\sigma_C=$3.369\,\AA,~and $\epsilon_C=$2.63\,meV.
For the substrate atoms we vary $\sigma$ in the range
(2.5\AA,~3.5\AA~) and $\epsilon$ in the range (10.0\,meV,
140.0\,meV), where the lower limits are typically for insulators,
e.g. SiO$_2$~\cite{MD2010} and the upper limits are typical for
metallic substrates, e.g. Na, K, etc~\cite{neeknanotech,ercok}.
Notice that the main difference between the two set of parameters is
the energy parameter ($\epsilon$) which is varied over more than one
order of magnitude.

 The total vdW energy stored between
GE with $N$ atoms and a substrate with $M$ atoms, can be written as
\begin{equation}\label{Eq1}
E^{A}_{vdW}={\sum^{N}_{i\,=\,1}}{\sum^M_{j=1}}u(r_{ij}),
\end{equation}
where $r_{ij}\,=\,|\textbf{r}_i -\textbf{r}_j|$ and $\textbf{r}_i$
refers to the position of the $i^{th}$ carbon atom of GE and
$\textbf{r}_j$ refers to the $j^{th}$ atom of the substrate. Often
in MD simulations, one approximates the above sums by including only
the nearest neighbor atoms in order to reduce the number of
interactions. Such an approach is accurate in the case of
short-range potentials. Regarding the cutoff distance appropriate
for the LJ potential $r_c=3\sigma$, only those substrate atoms
inside a sphere having radius $r_c$ with origin at the position of
the $i^{th}$ atom of GE, interact most strongly with the $i^{th}$
atom, while outside this sphere, the interaction strength decreases
very fast. Therefore, in practice for each `$i$', the sum over $M$
can be truncated and limited to the atoms inside a sphere with
radius $r_c$. This is done by employing a neighbor list in our MD
simulation. In our study, the number of GE atoms is $N=$14400 (which
is equivalent to a graphene sheet with dimension $l_x=$19.17\,nm and
$l_y=$19.67\,nm) and the number of substrate atoms is $M=$6000 (only
in the case of the Gaussian bump we performed a simulation with
$N$=72000 and $M$=35000).

The adhesion energy can be obtained using ab-initio calculations and
can be estimated using classical models (e.g. LJ potential). The
present day patterns of deformation of large scale GE on top of
substrates is beyond reach of traditional ab-initio methods. Our
classical vdW model as based on the LJ potential are able to
simulate the realistic sizes and gives a vdW energy (the main term
in the binding energy) between two non-bonded systems. Note that
several ab-initio calculations have demonstrated that the vdW
interaction between nanoscale objects can be well approximated by a
LJ potential~\cite{PRL2010BN,nanotechnology2002,surfacescience}.

The gradient of the total potential energy, i.e.
$E_{total}=E_P+E^{A}_{bind}$, is the force experienced by the
$i^{th}$ carbon atom, $\textbf{F}_i=-\nabla_i E_{total}$ . In common
molecular dynamics simulations Newton's second law should be solved
numerically in order to determine the path of motion of the atoms.
In this study, the equations of motion were integrated using a
velocity-Verlet algorithm with a time step of 0.5~fs and the
temperature was held constant at $T$=10\,K by a Nos´e-Hoover
thermostat.

The atomic stress experienced by each $i^{th}$ atom can be expressed
as \cite{PRB2004,carbon}
\begin{equation}
\eta^{i}_{\mu \nu}=\frac{1}{\Omega}\left(\frac{1}{2}m v^{i}_{\mu}
v^{i}_{\nu}+\sum_{j \neq i} r^{\nu}_{ij} F^{\mu}_{ij}\right),
\end{equation}
where the inner summation is over all the carbon atoms which are
neighbors of the $i^{th}$ atom which occupies a volume $\Omega= 4\pi
a_0^3/3$. The quantities $m$ and $v^i$ denote the mass and velocity
of the $i^{th}$ atom  and the scaler $r^{\nu}_{ij}$ is the $\nu$
component of the distance between atoms `i' and `j'. $F^{\mu}_{ij}$
is the force on the $i^{th}$ atom due to the $j^{th}$ atom  in the
$\mu$ direction. We have used this expression to calculate the
stress on each atom. In order to be able to visualize the stress
distribution on the GE atoms, we colored the atoms according to the
value of the dimensionless stresses, i.e green (red) is related to
the minimum (maximum) possible stress.

\section{Continuum approach}

Evaluation of the  vdW  contribution of the stored energy in the
deformed GE with average density $\Sigma_G$, due to the interaction
with a substrate with average density $\Sigma_S$, is obtained after
the integration of the vdW potential over both GE and the substrate
surfaces. Here, we present for comparative purposes a continuum
model for the stored vdW energy between the GE membrane and the
substrate. Such an approach can be used to calculatie the vdW energy
stored between two objects~\cite{ACSnano,Langmuir2010,langmuir}.

In the absence of external pressure, the total free energy of a
membrane consists of three terms, i.e.  bending, stretching and vdW
terms which are given by
\begin{equation}\label{EqFr}
\emph{f}=\iint dx dy [\tau (\nabla h_G(x,y))^2+\kappa (\nabla^2
h_G(x,y))^2]+E^{C}_{vdW},
\end{equation}
where $\tau$ and $\kappa$ are the surface tension and the bending
rigidity of the membrane, respectively. The two first terms in
Eq.~(\ref{EqFr}) are relevant to the bending and stretching energies
of  GE~\cite{Piniing} and the third term, $E^{C}_{vdW}$, is the
total vdW contribution of the interaction between the substrate and
the membrane. $E^{C}_{vdW}$ includes two repulsive and attractive
terms. The stored adhesion energy per area is determined by
\cite{langmuir}
\begin{equation}\label{EqF}
\chi_T=(\frac{\emph{f}_{min}}{A}-\tau),
\end{equation}
where $\emph{f}_{min}$ is the minimum of the total free energy when
the membrane takes its optimum configuration. `$A$' is the projected
area onto the $x-y$ plane, i.e. $A=\int dx dy$. Equation~(\ref{EqF})
was used by Swain \emph{et al}~\cite{langmuir,PREswin} to estimate
the adhesion energy of typical soft membranes over different
substrates. In Ref.~\cite{Piniing} the adhesion part in the free
energy  was taken as a  coupling constant. Assuming a planar local
relative height coordinate function for the vdW interaction energy
between the substrate and the soft membrane, i.e. the Deryagin
approximation, the vdW energy is approximated by
\begin{equation}\label{Eq.Dery}
E^{C}_{vdW}=\iint dx dy [V_0+\psi({\delta h}^2)],
\end{equation}
where $\psi$ is a function of the height increment $\delta
h=h_G(x,y)-h_S(x,y)$  and $V_0$ is a constant. Substituting
Eq.~(\ref{Eq.Dery}) in Eq.~(\ref{EqFr}) and minimizing the total
free energy with respect to $h_G(x,y)$ results in the following
differential equation~\cite{langmuir,PREswin}
\begin{equation}\label{EqDif1}
(\kappa\nabla^4-\tau\nabla^2+v)h_G(x,y)=v h_S(x,y),
\end{equation}
where $v$ is proportional to the Hamaker constant
($\propto\epsilon\sigma^6\Sigma_S\Sigma_G$). Equation (\ref{EqDif1})
can be solved in Fourier space~\cite{langmuir,PREswin}
\begin{equation}\label{EqDif2}
h_G(\textbf{k})=\frac{h_S(\textbf{k})}{1+k^2\xi_{\tau}+k^4\xi_{\kappa}},
\end{equation}
where $k$ is the wave vector, $\xi_{\tau}=\tau/v$ and
$\xi_{\kappa}=\kappa/v$.  Equation~(\ref{EqDif2}) was used to find
the optimum configuration of a soft membrane on top of corrugated
substrates~\cite{langmuir,PREswin}.

 Here we assume that the vdW energy is not localized and use the LJ
potential for the interaction between graphene and the substrate.
This gives us the vdW contribution to the adhesion energy, i.e.
$\chi$. We assume that both substrate and membrane are homogenous
and continuous materials. The obvious difference between the
atomistic model and the continuum model is the absence of the
chirality effect. The LJ potential energy between the GE membrane
and the continuum substrate is given by
\begin{equation}\label{EqC}
E^{C}_{vdW}=\iint_{G,S} \zeta(\textbf{r}_1)\zeta(\textbf{r}_2)
u(\Delta r_{12})\,ds_1 ds_2,
\end{equation}
where $\Delta r_{12}=|\textbf{r}_1 -\textbf{r}_2|$ and
$\zeta(\textbf{r}_1)=\Sigma_G(\textbf{r}_1)g_G(\textbf{r}_1)$ and
$\zeta(\textbf{r}_2)=\Sigma_S(\textbf{r}_2)g_S(\textbf{r}_2)$. Here,
$\Sigma_G=2/|\textbf{a}_1\times \textbf{a}_2|$ and $\Sigma_S$ are
the mean surface density of carbon atoms in the GE and  the
substrate lattice, respectively. The substrate density is
$\Sigma_S=\ell^{-2}$ which is equivalent to a (100) surface of a
crystal with lattice parameter equal to $\ell$.~The scaler
$g(\textbf{r}_i)=\sqrt{1+(\overrightarrow{\nabla} h(x_i,y_i))^2}$ is
defined by the appropriate transformation of a surface element in
curvilinear coordinates into a two-dimensional planar surface ($x-y$
plane in cartesian coordinate), i.e. the determinant of the metric
tensor . In Eq.~(\ref{EqC}) the position vector
$\textbf{r}_i=(x_i,y_i,h(x_i,y_i))$ refers to the position of the
surface element $ds_i$ of the GE membrane ($i=1$) or substrate
($i=2$). As already mentioned, because of the short range nature of
the LJ potential, we assume that the main contribution of the vdW
energy is due to the interaction with the outer surface of the
substrate. This model gives a good insight into the vdW adhesion
energy between GE and various substrates. For the continuum model we
will give only the results for GEs on top of a Gaussian
bump/depression. Notice that it is analytically impossible to
minimize Eq.~(\ref{EqFr}) by substituting $E^{C}_{vdW}$
(Eq.~(\ref{EqC})). Therefore, we used the optimum configuration for
the GE membrane as obtained from our MD simulations.

\section{Results and discussion}
In this study we investigate several different geometries for the
substrate which can be realized experimentally. The substrate atoms
are assumed to be rigid during the simulations which is a reasonable
approximation due to the different atomic-vibrations time scale  in
graphene and the substrate. In order to model the substrate, a (100)
surface having lattice parameter $\ell$=3\,\AA ~is used which is a
typical   lattice parameter. The density of sites in the substrate
is $\Sigma_S=\ell^{-2}$. Since the interaction between the substrate
and graphene is  weak and of short range (i.e. van der Waals
interaction) the main contribution to the mutual stored energy and
to the force between graphene and the substrate comes from  the
upper layer of the substrate. We can show this explicitly  by
calculating the energy and force as function of the number of atomic
layers in the substrate for one of our samples (We took the case of
graphene over a Gaussian bump see Fig. 12(a)). The  total energy can
be written as
\begin{equation}\label{test}
E^{A}_{vdW}(L)=\sum^{L}_{n=1}\sum^{N,M}_{i,j} u_n(r_{ij}),
\end{equation}
 where $u_n$ is the contribution of the n$^{th}$ layer and $L$ is the number
 of considered substrate layers. For instance for graphene
 on top of a Gaussian bump the attractive and repulsive parts are
  proportional to $(\rho^2+(z+n \ell)^2)^{-3}$ and $(\rho^2+(z+n\ell)^2)^{-6}$, respectively (here $\rho$ ($z$) is the
  planar (vertical) distance between an atom in graphene with one in the top layer of the substrate)
   which decrease fast with $n$. Recalling the discussion below Eq.~(3) we found that
   Eq.~(\ref{test}) for the considered system around the central points ($r<$30\,\AA)~the top layer
($n=1$) contributes almost 99$\%$ of the total energy, the second
layer contributes 1$\%$ and the contribution of the other layers are
neglectible. This motivated us to restrict our study to the top
substrate layer which helps us considerably to minimize the CPU
time. Note that for larger $\Sigma_S$  and  smaller $\sigma_T$ the
contribution of the second layer will increase.

 The height of the graphene and the
substrate atoms at each point ($x,y$) are denoted by $h_G(x,y)$ and
$h_S(x,y)$, respectively. The calculations are done for  two
different boundary conditions: i) free boundary condition, and ii)
supported boundary conditions which prevents  in-plane movements for
two longitudinal ends of GE. The two atom rows at the longitudinal
ends were taken in the zig-zag direction (in most cases) and they
are allowed to move in the $z$-direction.

\subsection{ Sinusoidal substrates: Free boundary condition}
A sinusoidal deformation of the substrate along the $x$-direction is
given by
\begin{equation}\label{Eqsine}
h_S(x,y)=h_0 \sin(k x),
\end{equation}
where $k=2\pi/\lambda$ and the amplitude is $h_0$. We used different
wave lengths, i.e. $\lambda=$2,~3~and~4\,nm and two sets of $\sigma$
and $\epsilon$, i.e. (3.5\,\AA,~10.0\,meV) and
(3.4\,\AA,~100.0\,meV) with fixed $h_0=0.5$\,nm. At the start of the
simulation, we put a flat graphene sheet on top of this substrate at
$h_G(x,y)=h_S(x,y)+r_{min}$. We choose the $x$-direction to be the
arm-chair direction.
\begin{figure}
\begin{center}
\includegraphics[width=.95\linewidth]{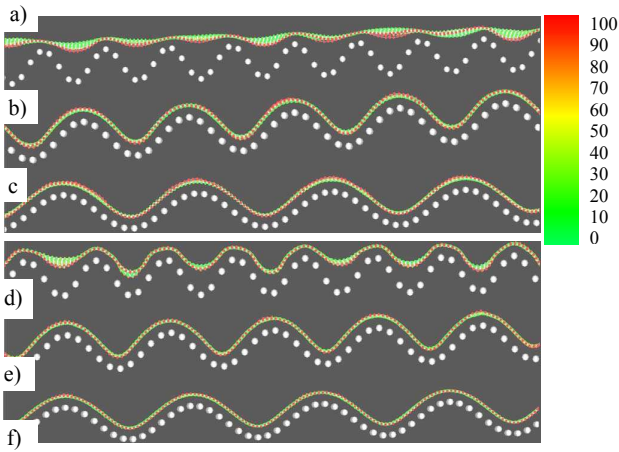}
\caption{(Color online)  Side views of the optimum configuration of
graphene on top of sinusoidal substrates with different wave
lengths. The filled white circles are the substrate atoms. The
parameters in (a,b,c) are $\sigma=3.5\,\AA,\epsilon=10.0\,meV$ and
in (d,e,f)  are $\sigma$=3.4\,\AA~and $\epsilon=100.0\,meV$. The
wave lengths are $\lambda=$2\,nm (a,d), $\lambda=$3\,nm (b,e) and
$\lambda=$4\,nm (c,f). For the graphene sheet, the colors represent
the stress distribution, the highest stress is denoted  by red and
the lowest
 by green.   \label{figsine1} }
\end{center}
\end{figure}

\begin{figure}
\begin{center}
\includegraphics[width=.95\linewidth]{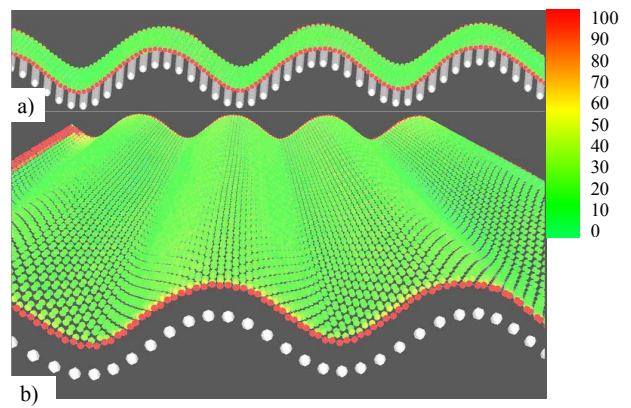}
\caption{(Color online) Two side views of Fig. \ref{figsine1}(c).
For the  graphene sheet, the colors indicate the stress
distribution. The substrate atoms are indicated by filled white
circles below graphene. The sinusoidal substrate is shown only by
the first front row of  atoms.  \label{figsin2} }
\end{center}
\end{figure}

 After 0.5\,ns of the MD simulation, GE found its optimum configuration which is deformed and corrugated.
 Figure~\ref{figsine1} shows six snap shots of free GE
(upper corrugated sheets in each panel) over three different
substrates (filled circles in each snap shot). In Figs.
\ref{figsine1}(a,b,c) the vdW parameters were set to
$\sigma$=3.5\,\AA,~$\epsilon$=10.0\,meV and in Figs.
\ref{figsine1}(d,e,f) the vdW parameters were set to
$\sigma$=3.4\AA,\,$\epsilon$=100.0\,meV. The wave length in Figs.
\ref{figsine1}(a,d) is 2\,nm. Notice that for $\lambda$=2\,nm GE
does not follow the substrate, i.e. $h_G(x,y)\neq h_1+h_S(x,y)$,
where $h_1$ is a vertical shift of the order of graphite's layer
distance, i.e. 3.4\,\AA.~By increasing the wave length or
$\epsilon$, GE follows more closely the substrate profile, i.e.
$h_G(x,y)\cong h_1+h_S(x,y)$. Increasing  $\epsilon$ yields stronger
adhesion and deforms GE (Fig.~\ref{figsin2} shows  zoomed versions
of Fig.~\ref{figsine1}(c)). Therefore, according to our MD
simulations, the shortest wave length which makes GE's profile
similar to the substrate's profile is larger than $\lambda=$2\,nm.
In all figures the colors indicate the stress distribution. Notice
that at the boundaries we always have red colors (i.e. maximum
stress) because of the presence of dangling bonds. In the other
parts, the distribution of stress is uniform (green colors). Notice
that in each particular system we scaled the colors by the highest
stress.

Here we compare our results to those predicted by continuum
elasticity theory for a membrane on top of sinusoidal surfaces. The
possible solution of Eq.~(\ref{EqDif1})~\cite{langmuir} for a
membrane on top of a deformation given by Eq.~(\ref{Eqsine}) is
\begin{equation}\label{EqHamaker}
h_G(x,y)=\frac{h_0\sin(kx)}{1+k^2\xi_{\tau}+k^4\xi_{\kappa}}.
\end{equation}
For  longer wave lengths, or small $\kappa$ and $\tau$ this solution
 gives $h_G(x,y)\approx h_S(x,y)$ which is in agreement with our MD
results for long wave lengths and large $\epsilon$. For graphene
membrane $\kappa\sim1.1$\,eV and $\tau \sim$\,eV/\AA$^2$~(taken of
the order of the Lam\'{e} coefficients) and Eq.~(\ref{EqHamaker}) is
valid if $v\gg \kappa, \tau$ which are related to both large
$\epsilon$, and $\Sigma_S$. The stiffer membrane with larger
$\kappa$ and $\tau$ can not curve easily and stronger adhesion due
to a larger $v$ ($\propto \epsilon$, i.e. stronger adhesion) is
required. By using
 $\sigma=(\sigma_C+\sigma_S)/2=(3.369+3.4)/2~\AA$,~$\epsilon=\sqrt{2.63\times10}~\,meV$, $\Sigma_{G}=0.225~\AA^{-2}$
 and $\Sigma_{S}=0.026~\AA^{-2}$ yields the Hamaker constant
$\approx4 \pi^2\epsilon \sigma^6 \Sigma_{G}\Sigma_{S}=1.78~\,eV$ and
consequently $v\sim 0.1 $ and thus  $k^2\xi_{\tau}\ll 1$ and
$k^4\xi_{\kappa} \ll 1$ yield $\lambda\gg4.7\AA$ and
$\lambda\gg5.6\AA$, respectively, which are in
 agreement with our MD finding, i.e. $\lambda > 20\AA$.

\subsection{ Sinusoidal substrates: Supported boundary condition}
\begin{figure}
\begin{center}
\includegraphics[width=1\linewidth]{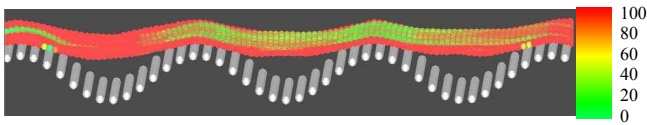}
\caption{(Color online)  The optimum configuration of graphene over
a substrate with  wave length $\lambda=4$\,nm. Graphene is supported
by the longitudinal ends while it can move along the $z$-direction.
Here $\epsilon=10\,meV$ and $\sigma=3.4\AA$~for the 12-6 LJ
potential (see also Fig. \ref{figsine1}(c) which shows graphene with
free boundaries over the same substrate). \label{figsinsup} }
\end{center}
\end{figure}

Here,  we impose the supported boundary condition on the
longitudinal ends (which is mostly taken to be the zig-zag
direction). In this case the atoms at $x=\pm l_x$ are not allowed to
move in-plane while they are allowed to move in the $z$-direction.
In this section we use the 12-6 LJ potential with parameters
$\epsilon=$10\,meV and $\sigma=$3.4\,\AA.~We repeated the above
simulations by applying the supported boundary condition along the
zig-zag direction for graphene on top of a substrate with
$\lambda=$4\,nm. Fig.~\ref{figsinsup} shows the optimum
configuration of the deposited graphene over this substrate after
minimization. Notice that the obtained  deformation is different
from those for free graphene over the same substrate,
Fig.~\ref{figsine1}(c). The reason is the supported boundary (i.e.
fixed in-plane) at the two longitudinal ends. Graphene does not
follow the substrate, but the lateral edges (along the $x$-direction
at $\pm l_y/2$) feel a much large stress as compared to
Fig.~\ref{figsine1}(c). Graphene is suspended across the periodic
peaks with a small curvature between them. The larger $\epsilon$
enhances the latter effect. Therefore, the vdW energy is not
dominating the bending energy of GE.

\subsection{ Step: Free boundary condition}
The second class of interesting substrate configurations   are steps
which were recently studied in an experiment to measure the
electronic and morphology of  deposited graphene~\cite{stepPRB}
\begin{equation}\label{Eqstep}
h_S(x,y)=h_0\theta(x),
\end{equation}
where $\theta(x)$ is the Heaviside step function with step height of
$h_0=$1\,nm and with $\Sigma_S$ density of sites. Both GE with
arm-chair and zig-zag direction are put on top of the steps.

\begin{figure}
\begin{center}
\includegraphics[width=0.95\linewidth]{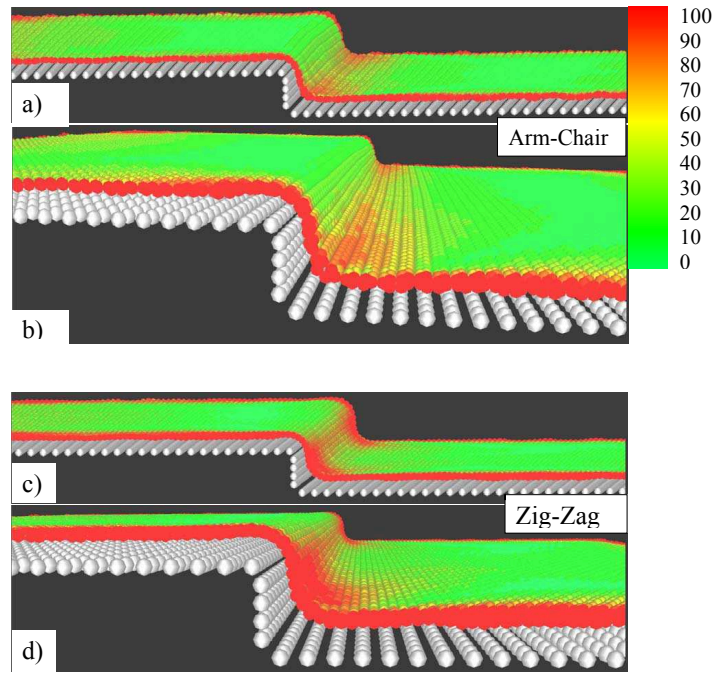}
\caption{(Color online) Arm-chair (a,b) and zig-zag (c,d) graphene
on top of two steps of height 1\,nm shown along two different angle
of view. The highest stressed atoms are shown by red and the lowest
by green. The substrate atoms are shown by filled circles  below
graphene. \label{figstep}}
\end{center}
\end{figure}

Figure~\ref{figstep} shows two snap shots of the optimum
configurations (an arm-chair GE with two different directions of
view in (a,b) and a zig-zag GE with two different direction of view
in (c,d)) of free GEs on top of steps. All GEs follow the steps
except around $x\approx0$ where GE is bent in a continuous  fashion
and does not follow the substrate. There are no considerable
differences between the optimum configurations of the free arm-chair
(Figs.~\ref{figstep}(a,b)) and free zig-zag
(Figs.~\ref{figstep}(c,d)) GEs on top of the step. Indeed, the wall
atoms at $x=0$ shift (due to adhesion) in both the left and the
right parts of the GE towards $x=0$.

\subsection{ Step: Supported boundary condition}
\begin{figure}
\begin{center}
\includegraphics[width=0.98\linewidth]{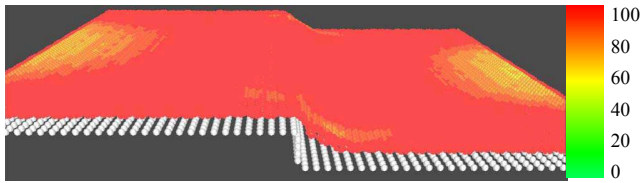}
\caption{(Color online)  The optimum configuration of arm-chair
graphene over a step located at $x=0$ with supported longitudinal
edges (see also Figs.~\ref{figstep}(a,b) which show  graphene with
free boundaries over the same substrate). Here $\epsilon=$10.0\,meV
and $\sigma=$3.4\,\AA~for the 12-6 LJ potential. \label{figstepsup}
}
\end{center}
\end{figure}

In Fig.~\ref{figstepsup}  the optimum configuration of GE along the
arm-chair direction with supported boundary condition is shown which
is over a sharp step defined by~Eq.~(\ref{Eqstep}). Notice that
there is  a significant difference between the deformation obtained
here and the one depicted in Fig~\ref{figstep}. The curvature around
the step ($x\approx0$) are different and all atoms of GE feel more
or less equal stress. The wall at $x=0$ adheres both the left and
the right part of the GE but the supported ends prevent fully
adhesion of GE to the wall, especially for the right hand side of
GE.

\subsection{ Trench: Free boundary condition}

\begin{figure}
\begin{center}
\includegraphics[width=0.9\linewidth]{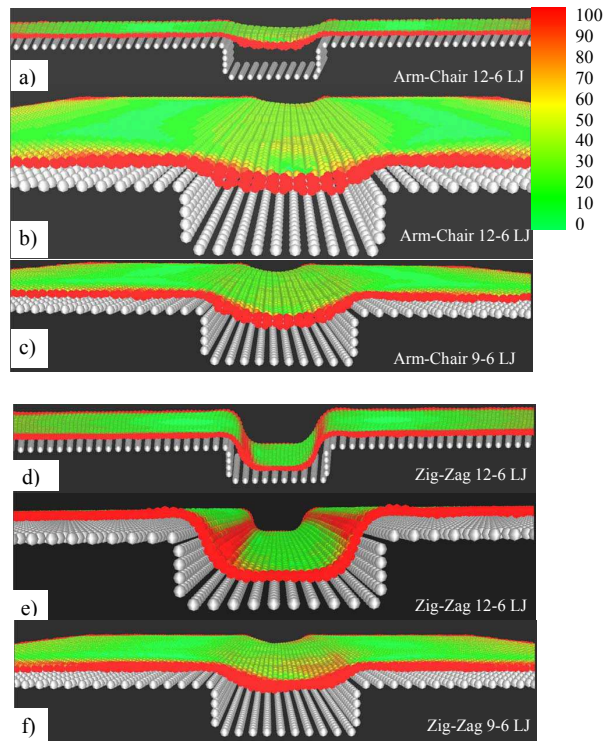}
\caption{(Color online)  Arm-chair (a,b,c) and zig-zag (d,e,f)
graphene on top of trenches  1\,nm deep and 3\,nm wide along the
$y$-direction. Panels (a),(b) (and also (d),(e)) are the same with
different angle of  view. For a graphene sheet, the colors indicate
the stress distribution. The substrate atoms are indicated by filled
circles below the GE. Zig-zag graphene follows the trench in
contrast to  arm-chair graphene. In (a,b,d,e) and (c,f) we used the
12-6 LJ and the 9-6 LJ potential, respectively. \label{figwell} }
\end{center}
\end{figure}

The other important  substrate that we studied here is an elongated
trench
\begin{equation}\label{Eq.trench}
 h_S(x,y)= h_0{\theta(x^2-d^2)},
\end{equation}
with two walls located at $x=\pm d=\pm1.5$\,nm with step height of
1\,nm and with $\Sigma_S$ density of sites. Figure~\ref{figwell}
shows two snap shots of GE on top of such trenches (an arm-chair GE
with two different angle of view in (a,b,c) and a zig-zag GE with
two different angle of view in (d,e,f)). After MD minimization
zig-zag GE follows the trench except around $x\approx \pm1$\,nm. In
this region, zig-zag GE is bent and does not follow the substrate.
There is a significant difference between the optimum configurations
of arm-chair (Figs.~\ref{figwell}(a,b)) and zig-zag
(Figs.~\ref{figwell}(d,e)) GEs. An arm-chair GE does not follow the
trench as well as a zig-zag GE. We attribute this effect to the
larger number of atoms of zig-zag GE (as compared to arm-chair GE)
in the well region ($|x|\leq d$), which yields a strong attractive
force on the GE atoms due to the substrate atoms within the well's
wall. Recently, Lu \emph{et al}~\cite{JAPL2010}  studied wider
trenches with $2d=$28.6\,nm using a 9-6 LJ potential in order to
find the vdW adhesion of GE membranes. In our study we also used the
9-6 LJ potential and found different deformations as compared to the
12-6 LJ potential, see Figs.~\ref{figwell}(c,f). This is due to the
different strength of both  attractive and repulsive parts in the
two models.

\subsection{Trench: Supported boundary conditions}
\begin{figure}
\begin{center}
\includegraphics[width=0.98\linewidth]{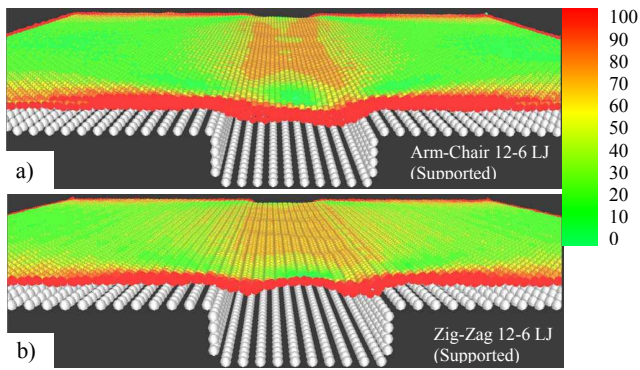}
\caption{(Color online)  The optimum configuration of arm-chair (a)
and zig-zag (b) graphene over two trenches located at $|x|<1.5$\,nm
where both  ends were supported in the $x-y$ plane (see also
Figs.~\ref{figwell}(a,d) which shows two graphene membranes with
free boundaries over the same substrate). Here $\epsilon$=10\,meV
and $\sigma=$3.4\,\AA~for 12-6 LJ potential. \label{figwellsup} }
\end{center}
\end{figure}

In Fig.~\ref{figwellsup} we show the optimum configuration of
arm-chair (a) and zig-zag (b) graphene with supported boundary
condition on top of the trench defined by~Eq.~(\ref{Eq.trench}).
There is a significant difference between the deformation obtained
here and those depicted in Fig.~\ref{figwell}. The curvature for
$|x|<d$ is very different and GE atoms around the well feel a lower
stress as compared to the one shown in Fig.~\ref{figwell}. Here both
arm-chair and zig-zag GE do not follow the substrate and were
suspended over the wells which is a consequence of the supported
boundaries. Therefore, by controlling the boundary condition one can
clearly control the GE deformation over the substrate.

\subsection{ Barriers: Free boundary condition}

\begin{figure}
\begin{center}
\includegraphics[width=0.9\linewidth]{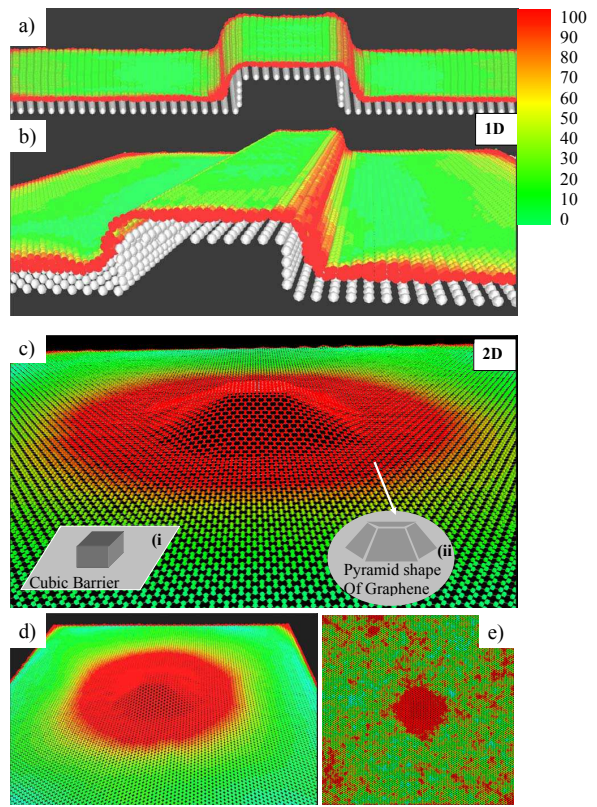}
\caption{(Color online) Elongated barrier (a,b)  and a cubic barrier
(c,d) deformation in the middle of a substrate covered with
graphene. The colors indicate the stress distribution in graphene.
The inset (i) in (c) shows a schematic model for the cubic barrier
and the inset (ii) is a schematic for  the optimum configuration of
graphene, i.e. pyramidal shape. Panel (d) is another view of (c).
The C-C bond lengths distribution for (c) (or (d)) are shown in (e),
where the red colors are related to the bonds around 1.47\AA~and
green colors are related to the bonds around 0.140\AA. Here
$\epsilon=10\,meV$ and $\sigma=3.4\AA$~for 12-6 LJ potential.
\label{figmbarrier} }
\end{center}
\end{figure}

A barrier in the middle of the substrate is the inverse situation of
the previous ones. An elongated barrier in the $y$-direction is
parameterized as
\begin{equation}\label{Eq.barrier}
h_S(x,y)=h_0{\theta(x^2-d^2)},
\end{equation}
with two walls at $x=\pm d=\pm$1.5\,nm with step height of 1\,nm and
with $\Sigma_S$ density of sites. Figs.~\ref{figmbarrier}(a,b) shows
two snap shots of arm-chair GE (with two different angle of view) on
top of the elongated barrier. As we see the stressed regions are
located around $x=\pm~d$. GE does not follow the rectangular shape
of the barrier in this part.

Another interesting case is the one of a substrate that consists of
a cubic barrier in the middle (see the inset (i) in
Fig.~\ref{figmbarrier}(c))
\begin{equation}\label{Eq.barrier2D}
h_S(x,y)= h_0{\theta(x^2-d^2)}{\theta(y^2-d^2)},
\end{equation}
with four walls at $x,y=\pm d=\pm1$\,nm with step height of 1\,nm
and with $\Sigma_S$ density of sites. Figs.~\ref{figmbarrier}(c,d)
show that the optimum configuration is
 pyramidal shaped (inset (ii) shows this schematically). This
particular deformation is due to the four corners of the cube which
strongly repel the GE.. The highest stresses are distributed around
the steps (red colors in the $|x^2+y^2|\leq 3d$). The C-C bond
lengths in GE are distributed non-uniformly
(Fig.~\ref{figmbarrier}(e)) but still around the barrier we have a
larger stretch in the bond lengths (up to $\simeq$ 0.147\,nm which
are shown by red colors).

\begin{figure}
\begin{center}
\includegraphics[width=0.98\linewidth]{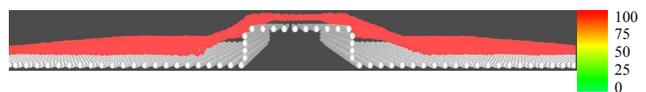}
\caption{(Color online)  The optimum configuration of arm-chair
graphene over an elongated barrier  of size $|x|<1.5$\,nm where
zig-zag edges were supported in the $x-y$ plane (see also
Fig.~\ref{figmbarrier} which shows graphene with free boundaries
over the same substrate). Here $\epsilon=10\,meV$ and
$\sigma=3.4\AA$~for 12-6 LJ potential. \label{figbarsup} }
\end{center}
\end{figure}

\subsection{ Barriers: Supported boundary conditions}
Fig.~\ref{figbarsup} shows the optimum configuration of arm-chair GE
in the case of  supported boundary condition over two different
barriers (defined by~Eqs.~(\ref{Eq.barrier}),~(\ref{Eq.barrier2D})).
As we see the stress distribution and the deformations are
completely different to those shown for free graphene (see
Fig~\ref{figmbarrier}).

\subsection{ Spherical bubble: Free boundary condition}
\begin{figure}
\begin{center}
\includegraphics[width=0.9\linewidth]{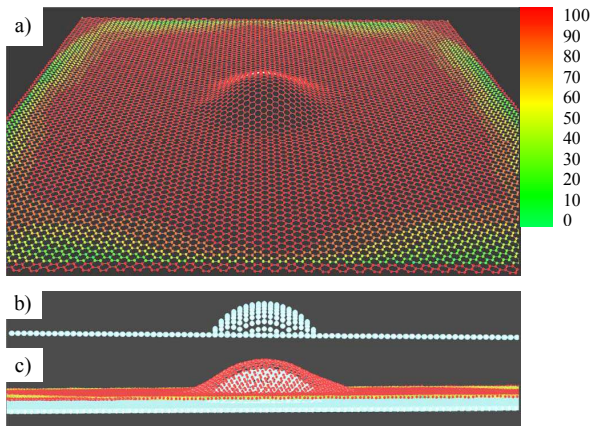}
\caption{(Color online) (a) The optimum configuration of the
graphene sheet over a bubble with $h_S(x,y)=\sqrt{R^2-\rho^2}-R/2$
deformation ($R=2\,nm$). The corresponding substrate is shown in
(b). In (c) we show both the GE and the substrate. For a graphene
sheet, the colors indicate the stress distribution. The highest
stress is shown by red color and the lowest by green. The substrate
atoms are indicated by filled white circles  below  graphene.
\label{figbubble} }
\end{center}
\end{figure}

The next important type of deformation for the substrate that has
been realized experimentally~\cite{Bubl1,Bubl2} is a bubble (see
Fig.~\ref{figbubble}(b)) which we model by
\begin{equation}
 h_S(x,y)=\sqrt{R^2-\rho^2}+h_1,\label{hbump}
 \end{equation}
where $R$ is the radius of the bubble and $\rho^2=x^2+y^2$. In order
to create an uniform discrete atomistic structure for the bubble, we
set $h_1=-R/2$. Figures~\ref{figbubble} (a,c) show the obtained
optimum configuration from MD simulation for the GE on top of a
bubble with $R$=2\,nm. The optimum configuration is a Gaussian. In
order to produce  uniform bubbles, we increased the density of
lattice sites in the bubble location where $\Sigma=1/4\,\AA^{-2}$.
Increasing the number of lattice sites on the bubble results in a
much larger stressed GE which influences regions far from the center
(see Fig.~\ref{figbubble}(a)).

\subsection{ Spherical bubble on the substrate: Supported boundary condition}

\begin{figure}
\begin{center}
\includegraphics[width=0.98\linewidth]{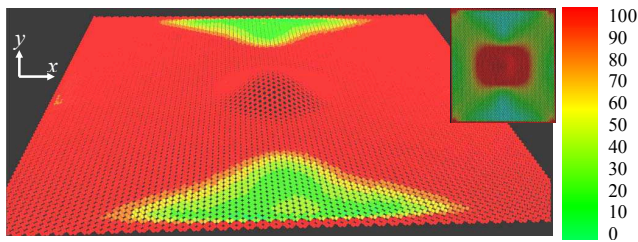}
\caption{(Color online)  The optimum configuration of arm-chair
graphene with two longitudinal ends  supported in the $x-y$ plane on
top of a bubble (see also Fig.~\ref{figbubble} which shows graphene
with free boundaries over the same substrate). The inset shows the
elongation of the deformation of graphene along the $x$-direction,
i.e. arm-chair direction. Here $\epsilon=10\,meV$ and
$\sigma=$3.4\AA~for 12-6 LJ potential. \label{figbublsup} }
\end{center}
\end{figure}

The optimum configuration for the supported graphene over a bubble
substrate is shown in Fig.~\ref{figbublsup}. Due to the supported
ends GE elongates longitudinally along the supported direction, see
inset in Fig.~\ref{figbublsup}.

\subsection{Gaussian bump/depression: Free boundary conditions}

There have been already a few studies that evaluated different
properties of a GE membrane in the presence of a Gaussian
deformation, but those studies did not address the following issues:
(i) the creation of the Gaussian deformation in GE using an
atomistic scale deformed substrate; (ii) what is the  effects of the
vdW energy strength on both the deformation and the adhesion energy
at the atomistic scale; (iii) what are the important differences
between the deformation due to a bump and due to a depression on the
atomistic scale, and (iv) what is the effect of the boundary
condition on GE.

\begin{figure}
\begin{center}
\includegraphics[width=1\linewidth]{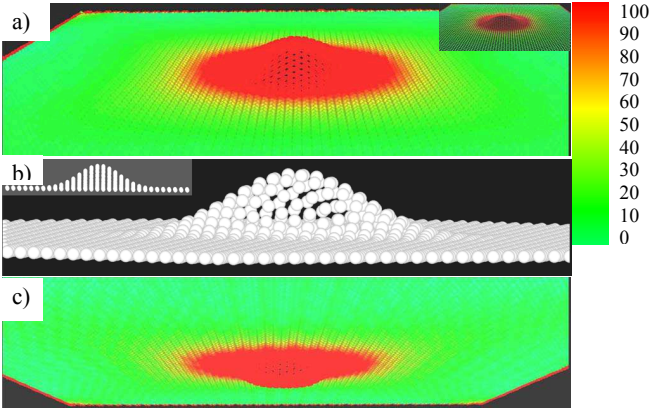}
\caption{(Color online) The optimum configurations obtained from
molecular dynamics simulations for a graphene sheet places on top of
a Gaussian bump (a) Gaussian depression (c) with
$h_S(x,y)=10\exp(-\rho^2/200)$(\AA).~ Here we see that graphene on
top of the Gaussian bump follows the substrate which is not the case
for a depression. For  a graphene sheet, the colors indicate the
stress distribution. Here $\epsilon=10\,meV$ and $\sigma=3.4\AA$.
Panel (b) is the bumped substrate, i.e.
$h_S(x,y)=10\exp(-\rho^2/200)$(\AA). \label{figbump} }
\end{center}
\end{figure}

The  Gaussian~bump (protrusion)/depression~\cite{neekJPCM,Piniing}
are parameterized as (Fig.~\ref{figbump}(b))
\begin{equation}
h_S(x,y)=\pm h_0 \exp(-\rho^2/2\gamma^2),\label{hbump}
 \end{equation}
 where $+h_0(-h_0)$ is the height (depth) of the Gaussian bump (depression)
and ${\rho}^2=x^2+y^2$ and $\gamma$ is the variance of the Gaussian.
Kusminskiy~\emph{et al} studied recently the pinning of GE over a
Gaussian bump in order to find the corresponding
attachment/deattachment of GE~\cite{Piniing}. Our model is more
realistic with relevant length scales for both height and variance
of the bumps/depressions. Figure~\ref{figbump}(a) shows a snap shot
of the optimum configuration of a GE on top of the Gaussian bump
(see Fig.~\ref{figbump}(b)) with $h_0=\gamma=1$\,nm located at the
center where $\epsilon=$10.0\,meV and $\sigma=$3.4\AA. The inset of
Fig.~\ref{figbump}(a) shows the far view of GE and the inset of
Fig.~\ref{figbump}(b) shows the side view of the Gaussian bump. The
red colors refer to the highest stresses which are mostly located
around the bump region, $r\leq 2\gamma$. For this size of the bump,
GE follows the Gaussian bump, i.e. $h_S(x,y)\approx h_1+h_G(x,y)$,
where $h_1$ is a vertical shift which is about graphite's layer
spacing 3.4\AA.

\begin{figure}
\begin{center}
\includegraphics[width=0.465\linewidth]{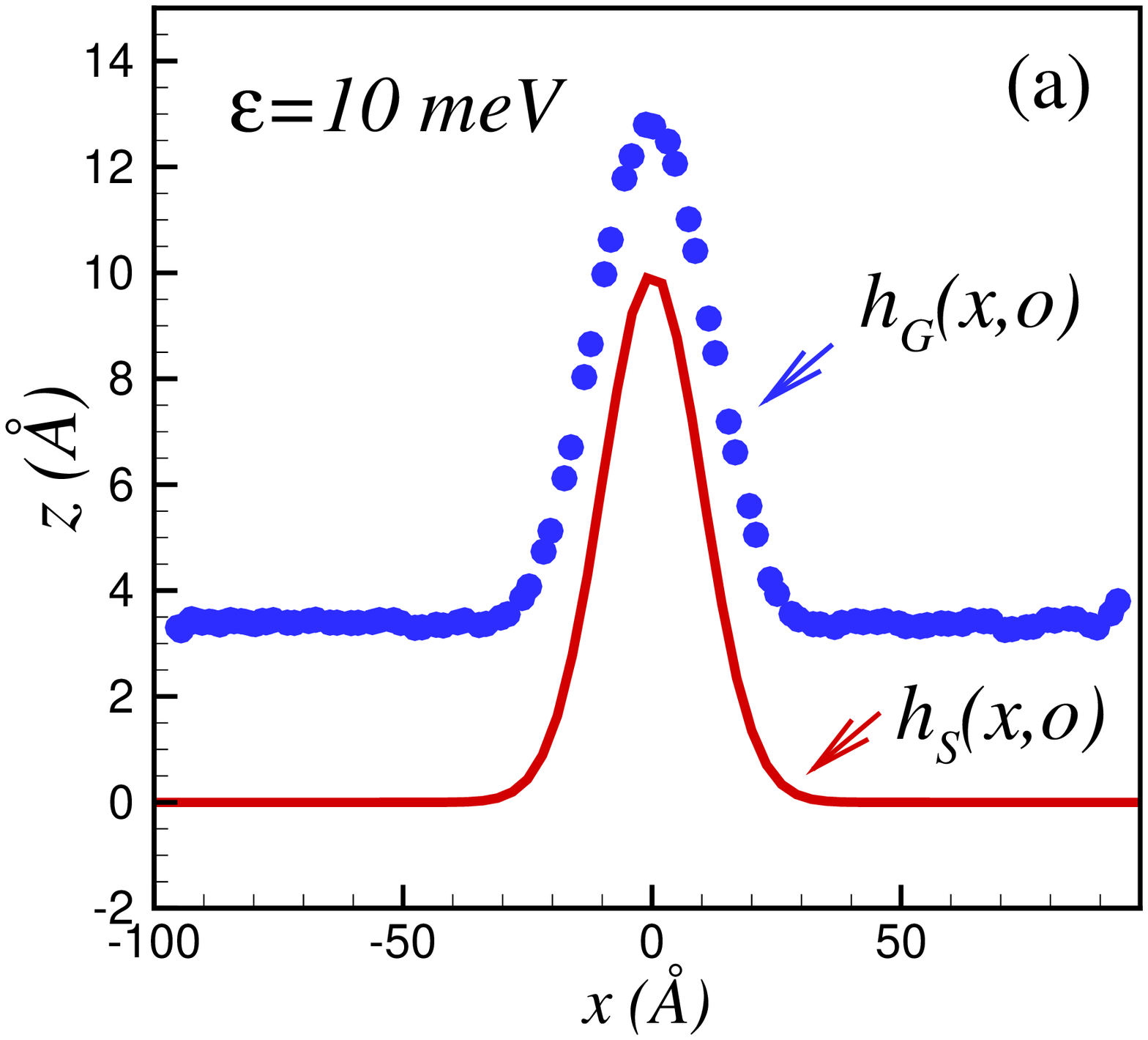}
\includegraphics[width=0.465\linewidth]{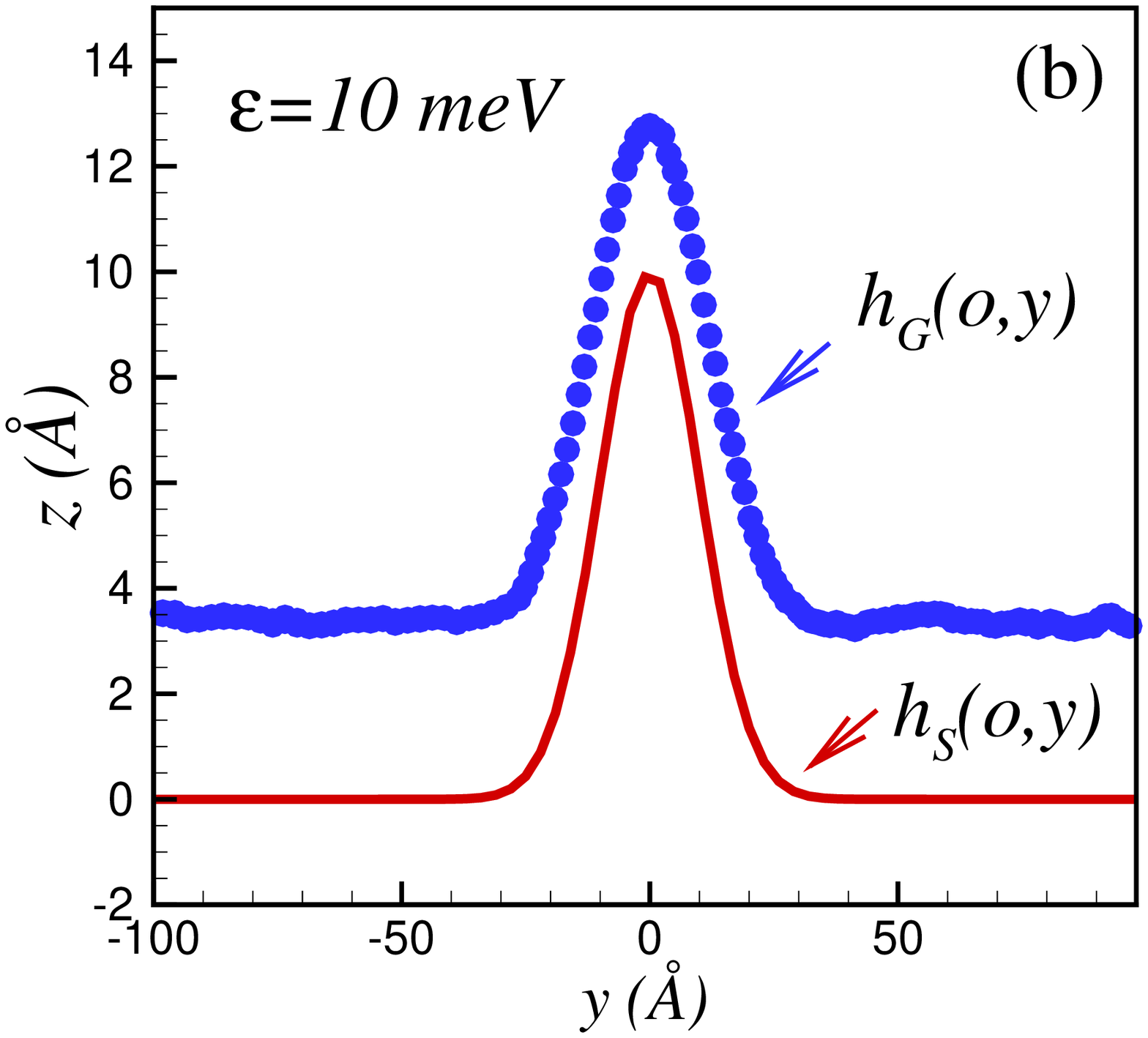}
\includegraphics[width=0.465\linewidth]{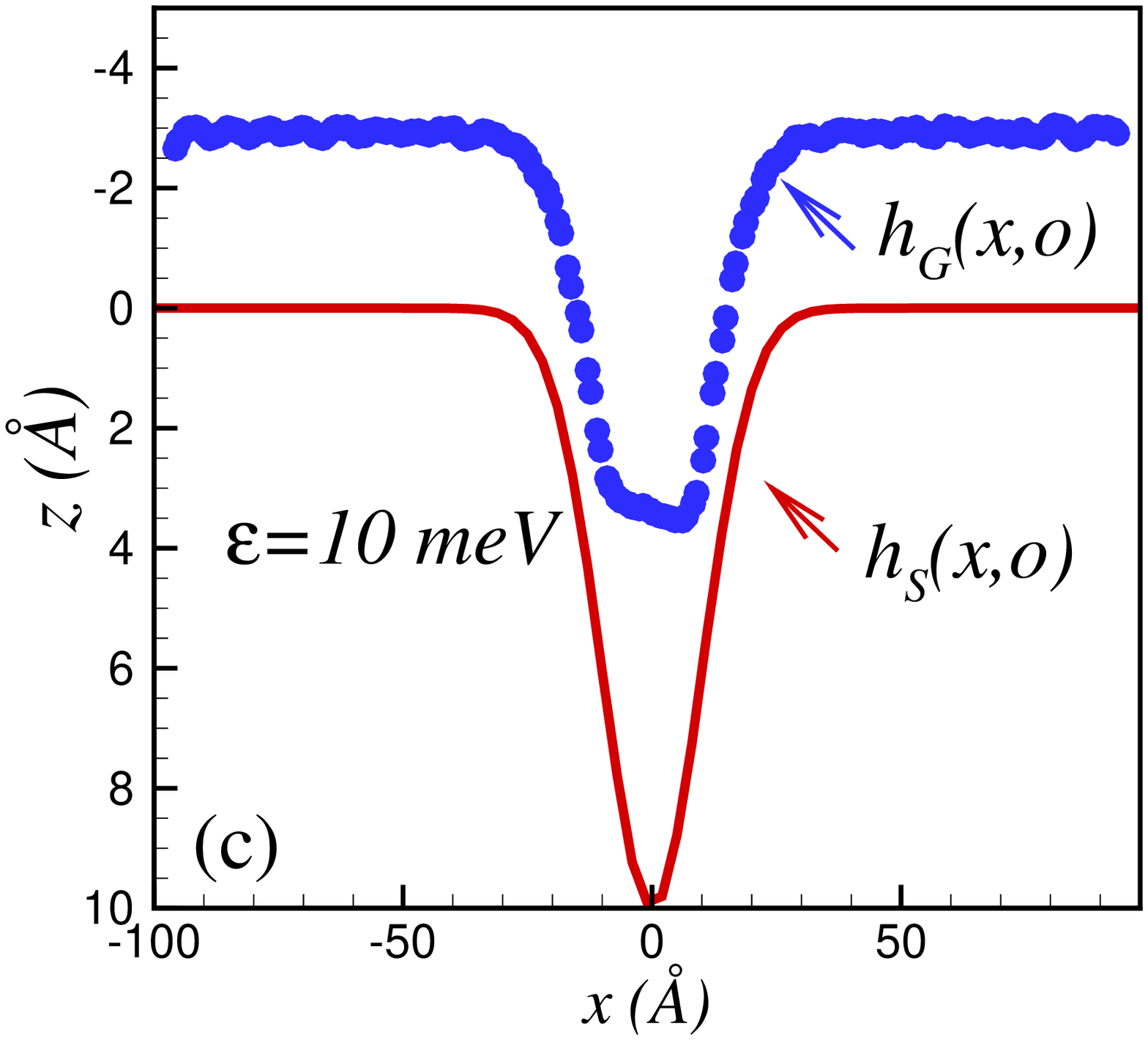}
\includegraphics[width=0.45\linewidth]{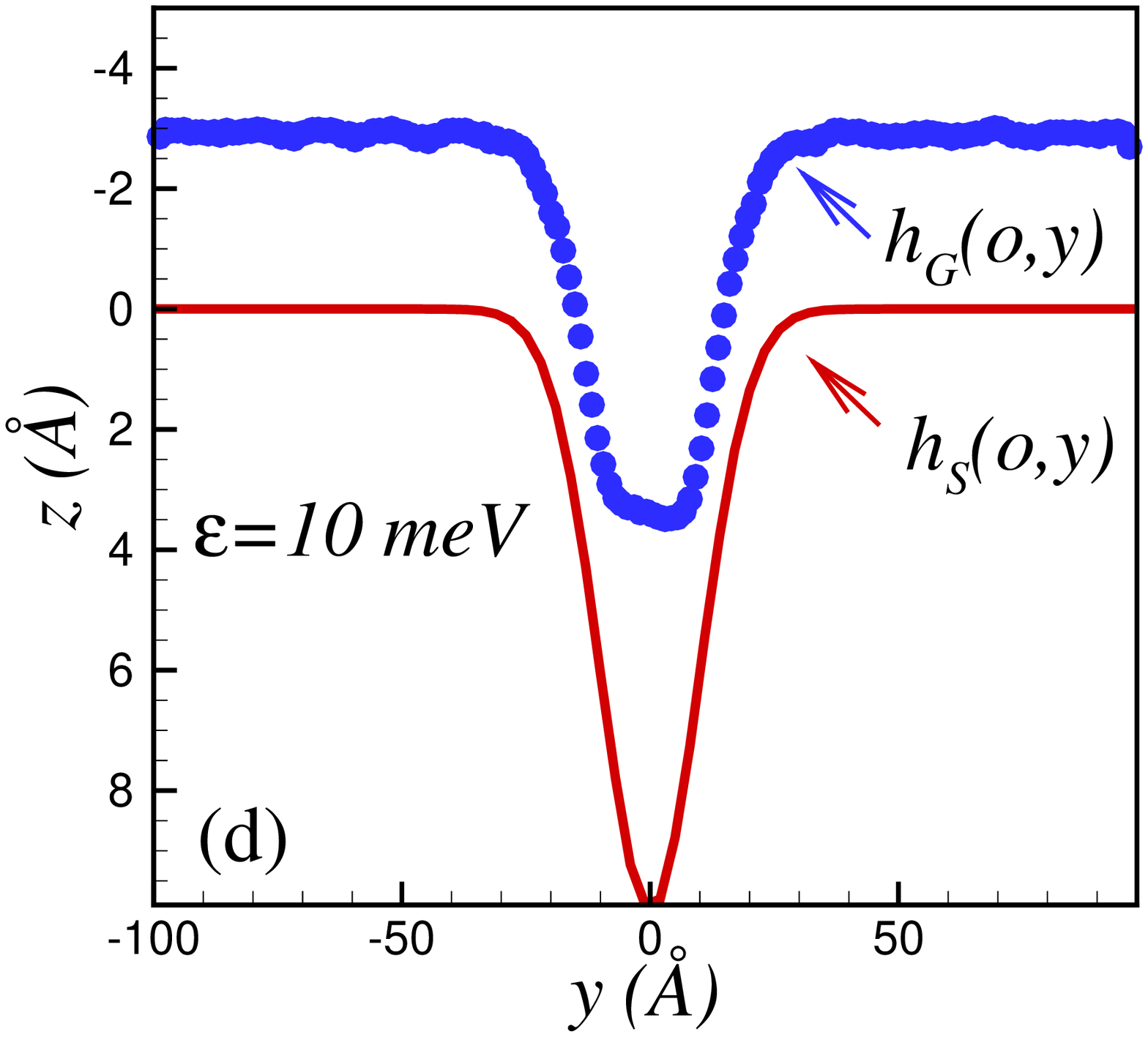}
\caption{(Color online) The heights of the substrate (red) and
graphene (blue) along $x=0$ and $y=0$ on top of a Gaussian
bump/Gaussian depression in $x$-direction (a)/(c) and $y$-direction
(b)/(d) (they are taken from the central portion of those
deformations shown in Fig.~\ref{figbump}). Circular points indicate
the C-atoms of graphene and the solid curves are the substrate.
Notice that graphene follows the bump which is not the case for a
depression. The colors indicate the stress distribution in the
graphene sheet,. Here $\epsilon=$10.0\,meV and $\sigma=$3.4\AA.
\label{figbump2} }
\end{center}
\end{figure}

Figure~\ref{figbump}(c) shows the optimum configuration of GE (found
from MD) on top of a depression with $h_0$=-1\,nm, and
$\gamma=1$\,nm. As we see the deformation of the GE over the
Gaussian bump is different from the one over the depression (while
both have the same variance, heights and potential parameters, i.e.
$\epsilon=$10.0\,meV and $\sigma=$3.4\AA). This is clear from the
curves shown in Fig.~\ref{figbump2} which were taken along $x=0$ and
$y=0$  (corresponding to the deformations shown in
Figs.~\ref{figbump}(a,c)). The optimum configuration of GE on top of
the depression (Figs.~\ref{figbump2}(c,d)) is not a Gaussian, i.e.
$h_S(x,y)\neq h_1+h_G(x,y)$, because of the stronger repulsive force
 inside the depression due to the interaction between GE and the
 substrate.

\begin{figure}
\begin{center}
\includegraphics[width=0.9\linewidth]{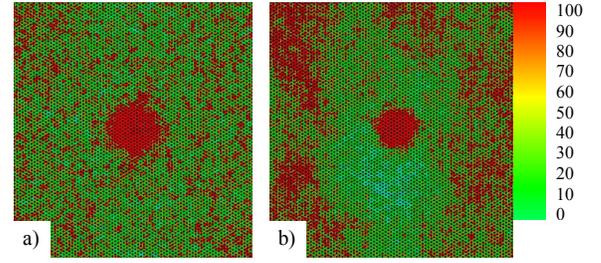}
\caption{(Color online) Bond length distribution in graphene on top
of a Gaussian bump (a) and a Gaussian depression (b). The longer
bond lengths are shown by red colors and the shorter bond lengths
are shown by green colors. \label{figmodel} }
\end{center}
\end{figure}

\begin{figure*}
\begin{center}
\includegraphics[width=0.465\linewidth]{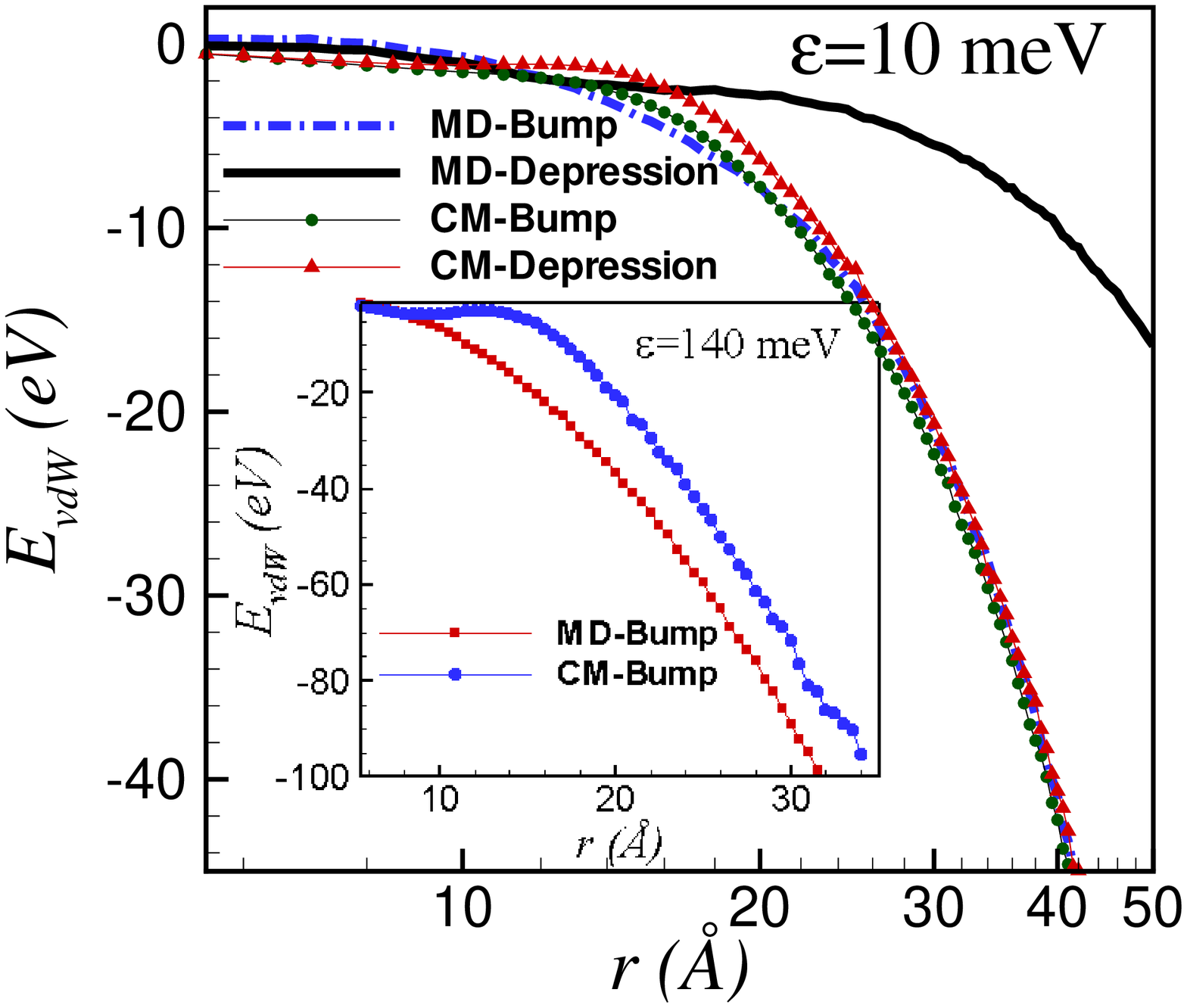}
\includegraphics[width=0.465\linewidth]{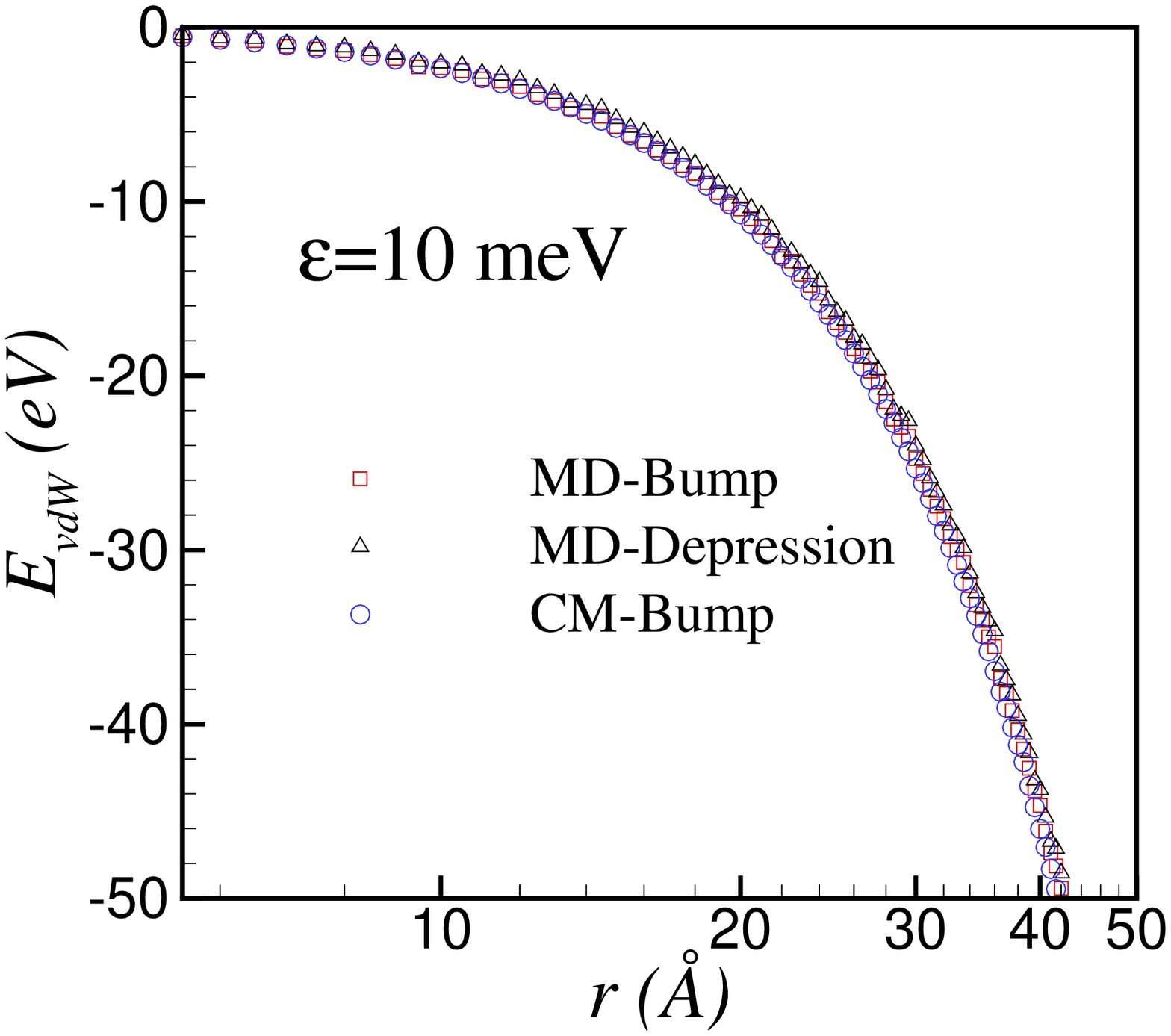}
\caption{(Color online)  (a) Variation of the van der Waals energy
versus the radius (in log-scale) measured from the origin (0,0).
Results are presented for  both molecular dynamics simulation (MD,
Eq.~(\ref{Eq1})) and continuum model (CM, Eq.~(\ref{EqC})). Here the
energy parameter in the LJ potential was set to 10\,meV (in the
inset we took 140\,meV) and $\sigma=3.4\AA$. The substrate is a
Gaussian bump with $10\exp(-\rho^2/200)$(\AA)~ deformation. (b)
Variation of the vdW energy versus the radius ($r$) for wider
Gaussian bump/depression with $10\exp(-\rho^2/900)$(\AA)~
deformation with $\epsilon=$10\,meV and $\sigma=3.4\AA$.
\label{figEAr10} }
\end{center}
\end{figure*}

Figure \ref{figEAr10} shows the variation of both $E^{A}_{vdW}$ (MD)
 (Eq.~(\ref{Eq1}))  and $E^{C}_{vdW}$ (CM) (Eq.~(\ref{EqC})) versus the radius, $r$, where $r$ is the upper limit
of the integrals in Eq.~(\ref{EqC}). In Fig. \ref{figEAr10}(a),  we
set $\gamma=h_0$=1\,nm, $\epsilon=10\,meV$ and $\sigma=3.4\AA$~which
are close to the one for the SiO$_2$ substrate \cite{MD2010}. For
the substrate with a  bump the energies of the atomistic model (MD)
are close to the one obtained from the continuum model (CM). For the
depression, our MD results are different  from the continuum model
results, which is a consequence of the different profiles in GE and
the substrate (see Fig.~\ref{figbump2}). Notice that the used
profile in CM for graphene on top of a depression is a Gaussian
profile (compatible to the substrate profile) while the found
optimized profile in the MD simulation as seen from Figs.~13(c,d) is
not a Gaussian. Therefore, there is a significant deviation from the
CM and MD results for depression as shown in Fig.~15(a). In the
inset of Fig.~\ref{figEAr10}(a), which is related to graphene on top
of a bump, we set the energy parameter to $\epsilon=140\,meV$. When
the
 energy parameter is large, the results of MD
and CM deviate, which is related to the strong attraction between
the substrate and GE.

Figure \ref{figEAr10}(b) shows the vdW energy for a Gaussian
bump/depression with larger variance, i.e. $\gamma=$3\,nm and
$h_0$=1\,nm, $\epsilon=$10\,meV and $\sigma=$3.4\AA. There is good
agreement between the results of MD simulations and those found from
the continuum model (CM). We conclude that for large variance the
continuum model provides a vdW contribution to the adhesion energy
which are comparable to the MD atomistic results. But for small
bump/depression the CM model is not applicable and the lattice
structure of GE should be taken into account.

\begin{figure*}
\begin{center}
\includegraphics[width=0.48\linewidth]{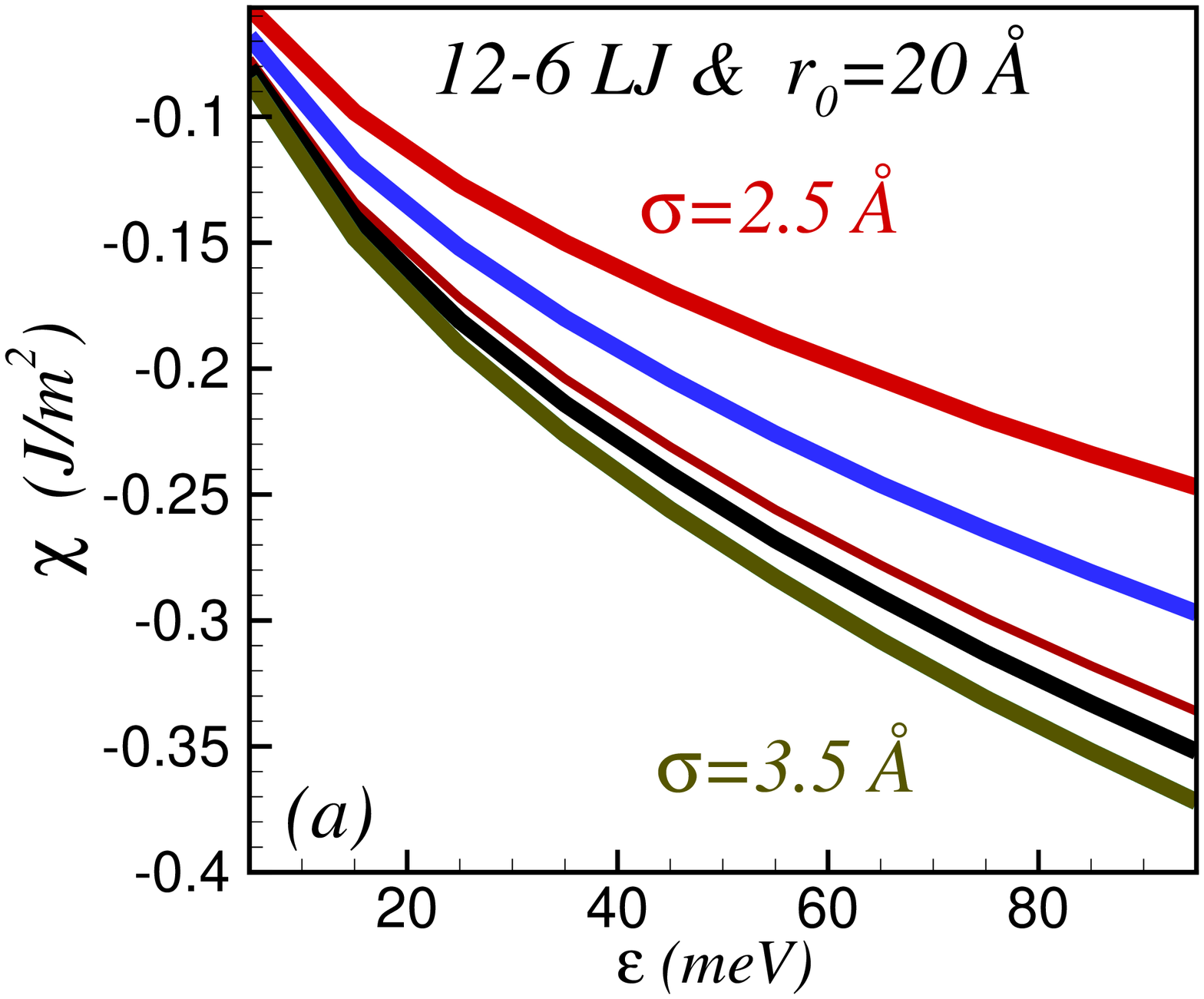}
\includegraphics[width=0.48\linewidth]{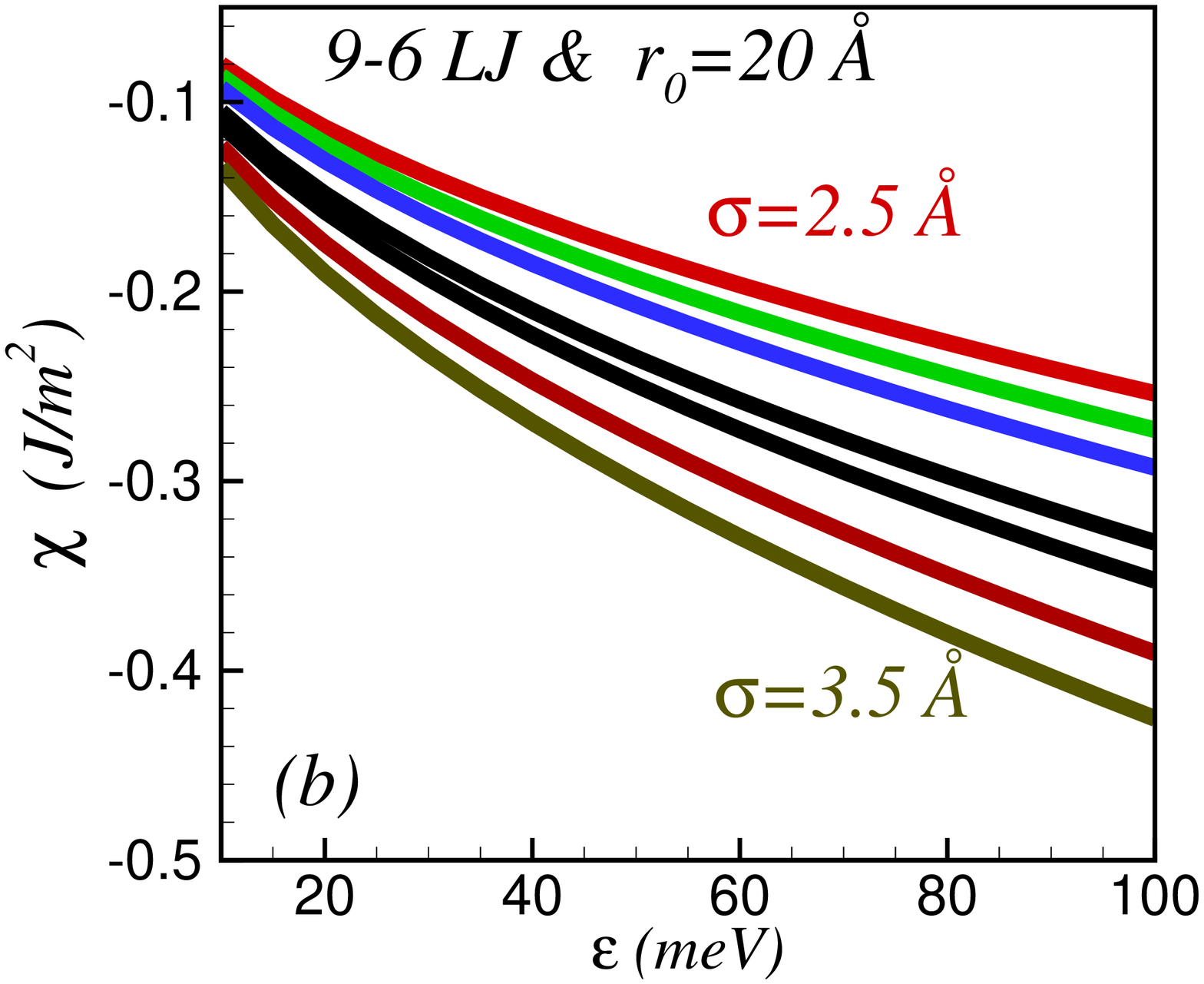}
\caption{(Color online)  Variation of van der Waals energy per area
versus $\epsilon$ from continuum model (Eq. (\ref{EqC})) for
graphene on top of a Gaussian bump with $10\exp(-\rho^2/200)$
deformation, where $\sigma$ increases from top to bottom with steps
of 0.25~\AA.~In (a),(b) we used 12-6 and 9-6 LJ potential
parameters, respectively. In both panels $r=r_0=2$\,nm which is the
upper limit of the integrals in Eq.~(\ref{EqC}) (see the text).
\label{figgamma} }
\end{center}
\end{figure*}

In Fig.~\ref{figgamma} the variations of  $\chi=E^{C}_{vdW}/A$
versus $\epsilon$ for various $\sigma$ (=2.5\,\AA,~3.5\,\AA)  are
shown. Here, graphene is deposited on top of a Gaussian bump with
$\gamma=$1\,nm, and $h_0$=1\,nm where $r=r_0=$2\,nm and the area is
calculated using $A=\pi r_0^2=4\pi\,nm^2$. In Fig.~
\ref{figgamma}(a) and Fig.~\ref{figgamma}(b) we used  the 12-6 LJ
and the 9-6 LJ potential parameters, respectively. The 9-6 LJ
potential gives results that have  larger $|\chi|$ for a particular
$\epsilon$ (Notice that in this paper the 9-6 LJ potential is used
only for  comparative purposes).

The energy per area, i.e. $\chi$,  is in the range of the adhesion
energy found for a graphene membrane positioned on top of SiO$_2$
substrate~\cite{naturenano}, i.e. 0.31-0.45\,J/m$^2$. However, note
that our results give  only the vdW contribution of the adhesion
energy which results from the standard $r^{-6}$ dispersion
interaction. 

\subsection{Gaussian bump/depression: Supported boundary conditions}
Since the optimum configuration of  supported graphene over a
Gaussian bump is similar to the one for a spherical bubble,  we will
not report them here. For supported graphene over a Gaussian
depression the optimum configuration is not Gaussian, similarly as
for free graphene over a depression (we do not show the optimum
configuration here).

\section{Summary}
We  carried out several molecular dynamics simulations and studied
systematically the optimum configuration of  large scale graphene
deposited  on top of several differently shaped substrates. The
stress distribution in graphene shows that highly stressed atoms
are located around the deformed regions of the substrate.\\

For short wave length ($\leq 2$\,nm) graphene is suspended across
the neighbor peaks of the sinusoidal substrate and thus graphene
does not follow the substrate. A graphene sheet on top of a cubic
barrier shows an unexpected pyramidal shape. It is found that for
large Gaussian bump/depression the van der Waals contribution in the
adhesion energy are in agreement with the prediction of the
continuum model. The van der Waals adhesion energy per area for a
nanoscale Gaussian bump is found to be around 0.1-0.35 J/m$^2$
depending on the energy parameter of the model. \\


\pagebreak
 \emph{{\textbf{Acknowledgment}}}. We thank  L. Covaci and S. Costamagna for valuable
comments.  We acknowledge M. Zarenia, M. R. Masir and D. Nasr for
fruitful discussions. This work was supported by the Flemish Science
Foundation (FWO-Vl) and ESF EUROCORE program EuroGRAPHENE: CONGRAN.

\end{document}